\documentclass{rmaa}
\newcommand{\kms}{\mbox{$\,$km s$^{-1}$}}

\newcommand{\mum}{$\mu$m} 
\newcommand{\cc}{$\rm{cm}^{-3}$}
\newcommand{\lsol}{L$_{\odot}$}
\newcommand{\msol}{M$_{\odot}$}
\newcommand{\mug}{$\mu$G}

\title{Spectroscopy of Stellar Jets, Outflows, and Young Stellar Objects with
the {\it  Infrared Space Observatory}}

\author{Alberto Noriega-Crespo\altaffilmark{1}
  \affil{SIRTF Science Center, \\
         California Institute of Technology}}
\fulladdresses{
\item SIRTF Science Center, California Institute of Technology,
      IPAC 100-22, Pasadena, CA, 91125
      (alberto@ipac.caltech.edu)}

\shortauthor{A. Noriega-Crespo}
\shorttitle{ISO Spectroscopy of YSOs}

\SetVolume{000} \SetFirstPage{001} \SetYear{2000}
\ReceivedDate{2000 Dec 15} 
\AcceptedDate{2001 Jan 16} 

\resumen{El Observatorio Infrarrojo del Espacio (ISO) fu\'e
una misi\'on europea muy exitosa, que nos ha dado una
vista del Universo en el infrarrojo sin paralelo alguno, incluyendo
cientos de observaciones de regiones de formaci\'on de estrellas
y flujos bipolares. Tres de los
equipos instrumentales, a cargo de la c\'amara infrarroja (CAM) y los 
espectr\'ometros de longitudes de onda corta (SWS) y larga (LWS), 
usaron una fracci\'on importante de su tiempo garantizado en el estudio 
espectrosc\'opico de objetos y flujos estelares j\'ovenes.
Aqu\'{\i} resumir\'e brevemente algunos de los hallazgos principales,
en particular la detecci\'on de agua, las lineas rotacionales de H$_2$ y
la presencia de otras mol\'eculas m\'as complejas. Presentar\'e nuevos
resultados espectrosc\'opicos de los flujos HH 1-2, HH 7-11 y Cep E, y 
de sus fuentes.  Finalmente, discutir\'e algunas de las tendencias generales 
que se obtienen de estas observaciones, as{\'{i}} como su importancia para 
entender la emisi\'on de esto objetos usando modelos de choque J y C.
}

\abstract{The Infrared Space Observatory (ISO) was an extremely successful 
european space mission that gave us an unparallel view of the Universe 
in the infrared, and provided us with  hundreds of observations of
star forming regions and bipolar outflows. Three of the instrument teams, 
in charge of the infrared camera (CAM) and the two spectrometers at short 
and long wavelengths (SWS and LWS respectively), 
used a significant fraction of their guarantee time to study 
YSOs and outflows spectroscopically. In here, I will briefly review
some of their main findings, particularly the detection of water, H$_2$ 
rotational emission lines and the presence of other complex molecules.
I will present new spectroscopic results on HH 1-2, HH 7-11 and Cep E, 
and their sources. And finally, I will discuss some of the general trends 
derived from these observations and their relevance in understanding the 
emission from these objects using J and C shock models.}
      

\begin{document}

\maketitle

\section{Introduction}
\label{intro}

After years of planning and developing the Infrared Space Observatory (ISO) 
was launched in November of 1995, carrying four state-of-the-art instruments 
and new mid/far infrared detectors. Over the next two and half years ISO
provided us with a wealth of incredible observations and a revolutionary 
view of the infrared universe. It was the first time that
we could look at mid/far infrared wavelengths with such sensitivity and 
angular resolution, and of course, a great opportunity to advance our knowledge
on star forming regions and related objects. 
For the spectroscopic study of outflows, jets and their 
sources it meant the possibility: (1) to determine the physical conditions 
of the atomic/ionic/molecular gas, (2) to describe in more detail their 
shock/ionization structure, (3) to understand the overall energy budget, 
(4) to study the relationship between the sources and the 
emitted spectra from their outflows, and (5) to analyze the global
trends provided by a bigger sample of objects. 

\subsection{ISO Instruments}
\label{sec:iso}

 ISO had a 60cm mirror and four instruments that were kept
a temperatures of 2-8 K using a large cryo-vessel with over 2000 liters
of superfluid Helium (Leech \& Pollock 2000).
A camera (ISOCAM), a photometer (ISOPHOT) and two spectrometers
at short (SWS) and long (LWS) wavelengths were the four scientific instruments
aboard ISO. ISOCAM and ISOPHOT had both spectroscopic capabilities, CAM with
a Circular Variable Filter (CVF) mode, and PHOT with Phot-S mode. The
four instruments had a 3\arcmin~unvignetted field of view (FOV).
Some of the main characteristics of these instruments are the following.
ISOCAM was a two channel camera with two InSb 32$\times$32 arrays which
covered a wavelength range from approximately 2.3-5.1\mum~(short) and 
5.0 - 17.3\mum~(long). Each channel allowed four different magnifications 
(1.5, 3, 6 and 12\arcsec/pixel), and in the CVF mode had a spectral resolution
R = $\lambda\over \Delta \lambda$ = 40 (Sibenmorgen et al. 1999).
The Short Wavelength Spectrometer (SWS) covered the wavelength range
2.38-45\mum,  with spectral resolutions from R = 400 (low resolution grating
spectrum) to R = 20,000 (Fabry-Perot). It had different apertures depending 
on the wavelength region of the grating; at the short range (2.38-12\mum) 
14\arcsec$\times$20\arcsec; at the long range 14\arcsec$\times$27\arcsec~
(12-27.5\mum), 20\arcsec$\times$27\arcsec~(27.5-29\mum) 
and 20\arcsec$\times$33\arcsec~(29.0-45.2\mum).
At least 4 different types of detectors were used in the grating
mode: InSb (2.38-4.08\mum), Si:Ga (4.08-12.0\mum), Si:As (12.0-29\mum) and
Ge:Be (37-48\mum) (Leech, de Graauw et al. 2000).
The Long Wavelength Spectrometer (LWS) covered the wavelength range
43 - 196\mum,  with spectral resolutions from R = 200 (medium resolution
grating spectrum) to R = 10,000 (Fabry-Perot). It had a circular FOV 
of $\sim 80$\arcsec, and three different types of the detectors were used
GeBe (43-50.5\mum), Ge:Ga (49.5-110\mum) and stressed Ge:Ga (103-196\mum)
(Gry et al. 2000).

\subsection{Key Programs on Star Formation, Outflows, Jets and their Sources}
\label{sec:key}

ISOCAM, SWS and LWS dedicated a substantial fraction of
their guarantee time to carry out spectroscopic observations 
of jets, outflows and their sources.
The ISOCAM group led by Sylvie Cabrit obtained nearly 100 
observations in their  project to
study the mid-infrared emission associated with energetic bipolar outflows
and the circumstellar dust halos around YSOs.
One the scientific goals of this project was to detect the [Ne~11] 12.8\mum~
emission line, as an indicator of relatively high velocity shocks 
($\ge$ 60\kms), as well as to analyze the effect of the UV field created 
by such shocks on the nearby dust particles.

The programs of Pre-Main Sequence stellar evolution and the nature of YSOs
of the LWS consortium were led by Paolo Saraceno. These
programs obtained more than 200 observations, including several 
full grating SWS spectra as well, of objects as classical T Tauri,
Herbig Ae/Be and FU Orionis stars (Benedettini et al. 2000; 
Giannini et al. 1999; Lorenzetti et al. 1999, 2000; Nisini et al. 1999; 
Spinoglio et al. 2000).

The SWS team had several projects to study the nature and evolution of
interstellar dust, both in the dense environment associated with
molecular clouds and low mass protostars, as well as that related with
intermediate and  massive young stars. Some of the topics studied by
the researchers working with them (e.g. W. Van Dishoeck, D. Whittet, 
and P. Wasellius) explored different aspects of the gas-phase and solid (ices)
chemistry of species such as H$_2$O and CO$_2$, and the chemical state of
molecules like OH, C$_2$H$_2$, CH$_3$ and CH$_4$.
Again, these programs obtained nearly 200 observations, and has lead to the
publication of several articles, and at least two very complete PhD. Theses
by A. Boogert (``The Interplay between Dust, Gas, Ice and Protostars'') and
M. van den Ancker (``Circumstellar Material in Young Stellar Objects'').

Besides these GTO projects, textually hundreds of observations in the Open
Time were awarded to study problems on star formation and outflows.
All the ISO data became public
in December 1998, making available over 30,000 scientific observations, the
result of approximately 1000 individual programs.

\section{Thermal Water Emission}
\label{water}

\begin{figure}
  \begin{center}
    \leavevmode
     \includegraphics[height=8cm,width=6cm]{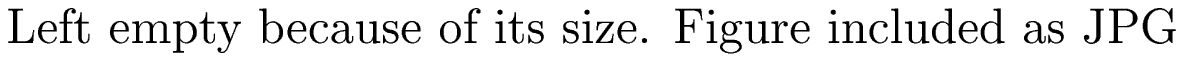}
    \caption{Fractional cooling in MHD molecular shocks (C-shocks)[from
    Draine et al. 1983]. At high densities and velocities
    H$_2$O, H$_2$ and [O~I] dominate the overall cooling.}
    \label{fig:anc_f01.ps}
  \end{center}
\end{figure}

\begin{figure}
  \begin{center}
    \leavevmode
     \includegraphics[height=8cm,width=6cm]{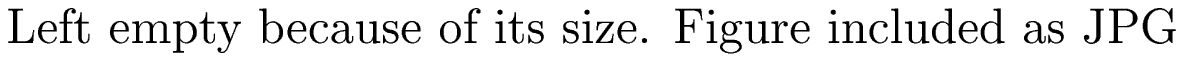}
    \caption{Emission spectra from a MHD molecular shock (C-shock) at
     40\kms, n(H$_2$) = $10^5$ \cc, B=447\mug~of the main molecular species
     from 1 - 2000\mum~[from Kaufman \& Neufeld 1996]}
    \label{fig:anc_f02.ps}
  \end{center}
\end{figure}

Among one of the most significant results obtained by ISO was the detection of 
water in active star forming regions. The theory already had predicted that in
molecular shocks or magneto-hydrodynamic C-type shocks (Fig. 1), at relatively
high densities ($10^4-10^6$\cc), there was a range in velocity  (15-40\kms) 
in which water was the most important coolant (Draine et al. 1983). Under such 
conditions the shocks eroded ice mantles from the dust grains, returning the 
H$_2$O to its gas phase and making available several transitions to release 
the energy in the far infrared. One of the most striking examples of this 
process detected by ISO was HH 54 (Liseau et al. 1996), where LWS found 20 
or more ortho and para-water emission lines and such that 
n($H_2$O)/n(H$_2$) = $10^{-5}$, i.~e. a lot higher than the expected 
value in quiescent molecular clouds ($10^{-7}$-$10^{-6}$), and with a cooling 
comparable to the mechanical energy of the outflow ($10^{-2}$\lsol). 

Another important example was that of Orion BN-KL object 
(Harwit et al. 1998), where LWS Fabry-Perot observations detected 8 different 
transitions, with an abundance n($H_2$O)/n(H$_2$) = $5\times 10^{-4}$, 
in agreement with the latest theoretical models of the emitted spectra 
(Fig. 2) by C-shocks (Kaufmann \& Neufeld 1996).

\section{HH 7-11 Outflow}
\label{hh711}

\begin{figure}
  \begin{center}
    \leavevmode
     \includegraphics[height=10cm,width=8cm]{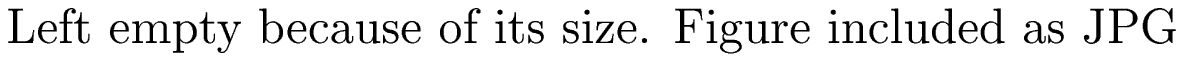}
    \caption{ISOCAM CVF image of HH 7-11 at H$_2$ (0,0) 6.91\mum
     (continuum subtracted). The emission from this flow in the mid-IR
     is dominated by H$_2$ pure rotational lines. 
     Spectra from individual knots are shown in detail. FOV$\sim$90\arcsec.}
    \label{fig:anc_f03.ps}
  \end{center}
\end{figure}

The bipolar outflow defined by the HH 7-11 system is one of the brightest and
best studied. Its redshifted outflow lobe is invisible at optical wavelengths,
and so the study of this object permits to compare the properties of embedded
outflows with those of optical HH objects.
We have analyzed recently the mid/far infrared physical characteristics
of HH 7-11 using SWS and LWS observations (Molinari et al.
2000). We found atomic ([O~I] 63\mum~and 145\mum, [Si~II] 34.8\mum) and 
molecular (H$_2$, CO and H$_2$O) emission lines along the bipolar flow.
Indeed, there was no significant difference in their properties
between the optical and the
invisible outflow component. We determine that emission could be explained
by a combination of J and C-shocks with velocities of $20-50$\kms~and 
preshock densities of $10^4-10^5$\cc. 
The SWS and LWS have a limited spatial resolution because of their apertures, 
and so the CVF ISOCAM observation of HH 7-11 provide further information
on its emission properties. Fig. 3 shows the outflow in the pure rotational 
line of H$_2$ S(5) emission at 6.91\mum~(a single data cube plane) together 
with the spectra obtained at each pixel, including HH 7, 8 and 10.
We notice that (i) the spectra are dominated by the H$_2$ pure rotational lines
S(7) through S(2), (ii) that the morphology of th system is almost identical
to that of  H$_2$ (1,0) S(1) emission at 2.12\mum, and (iii) behind HH 10
close to the position of HH 11, there is a trace of [Ne~II] 12.8\mum.
This presence of [Ne~II] indicates J-shocks with velocities $\ge 60$\kms,
a bit higher than that 40-50\kms~inferred from optical spectroscopic 
measurements (Solf \& B\"ohm 1990). The H$_2$ rotational lines are consistent
with our previous findings (Molinari et al. 2000) of the excitation 
temperatures ($\sim 500-800$K) and column densities 
($\sim 2-12\times 10^{19}$$cm^{-2}$) (Noriega-Crespo et al. 2001).

\section{Cepheus E Outflow}

Another example of an embedded/optical outflow is that of Cep E. A bright 
knot in the southern outflow lobe is visible in H$\alpha$ and [S~II] 6717/31, 
and known as HH 337. Some of the interesting characteristics of Cep E is that 
appears to be driven by an unique Class 0 protostar 
(Lefloch et al. 1996), and the outflow itself is very young, with a 
dynamical age of $\sim 3000-5000$ years. Cep E was observed with ISOCAM CVF 
and LWS in full grating mode, and we have recently analyzed these observations
(Moro-Martin et al. 2000).  Fig. 4 shows the CVF image in the S(5) (0,0) 
emission at 6.91\mum, with the spectrum at each pixel. We notice that both
outflow lobes are dominated by H$_2$ pure rotational emission lines, and 
{\it surprisingly} the IRAS 230111+6126 source is detected. Also that
this emission is remarkable similar to that at
2.12\mum~(Eisl\"offel et al. 1996; Ayala et al. 2000). In this case, 
we found T$_{ex}\sim 950-1300$K and N(H$_2$)$\sim 1-3\times 10^{19}$cm$^{-2}$,
i.~e. a bit higher temperatures and smaller column densities than in HH 7-11.

The LWS spectrum of the North and South outflow lobes had more than 20
molecular lines of CO, H$_2$O and OH (Fig. 5). From the CO, H$_2$O and H$_2$
coolings we determine that the FIR spectra was generated by C-shocks at
20-30\kms. The [O~I] 63\mum~and [C~II] 158\mum~fine structure emission lines
were also detected with similar strengths, indicating contamination by 
photodissociation in the [C~II] line. The corrected [O~I] 63\mum~collisionally
excited line required J-shocks at $\sim 20-30$\kms~to be explained.
Water was overabundant by factors of $10^2-10^3$, i.~e. consistent with 
Cep E low excitation and youth.

\begin{figure}
  \begin{center}
    \leavevmode
     \includegraphics[height=10cm,width=7.5cm]{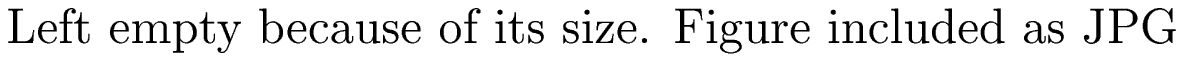}
    \caption{ISOCAM CVF image of Cep E at H$_2$ (0,0) 6.91\mum.
      The emission from this flow in the mid-IR
     is dominated by H$_2$ pure rotational lines. 
     Spectra from individual knots are shown in detail. FOV$\sim$90\arcsec.}
    \label{fig:anc_f04.ps}
  \end{center}
\end{figure}

\begin{figure}
  \begin{center}
    \leavevmode
     \includegraphics[height=7cm,width=8cm]{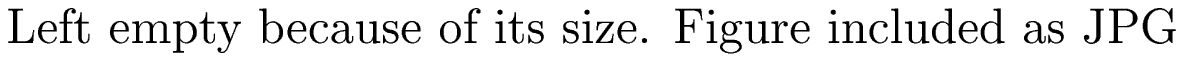}
    \caption{Full grating LWS spectra of Cep E North and Cepe E South.
     The spectra are rich in CO, H$_2$O and OH emission lines. [O~I] 63\mum~
     \& [C~II] 158\mum~are detected and equally strong.}
    \label{fig:anc_f05.ps}
  \end{center}
\end{figure}

\section{HH 1-2 Outflow}

\begin{figure}
  \begin{center}
    \leavevmode
     \includegraphics[height=7cm,width=8cm]{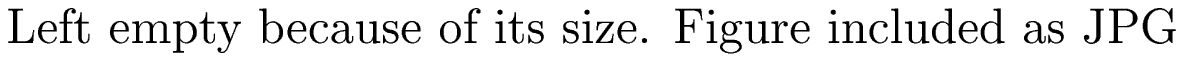}
    \caption{A comparison of HH 1-2 and Cep E outflows in the FIR to stress
     the difference between a high excitation optically visible object 
     (HH 1-2) and a low excitation embedded one (Cepe E).}
    \label{fig:anc_f06.ps}
  \end{center}
\end{figure}

The brightest and well studied HH 1-2 system, was also observed by ISO 
with SWS medium resolution, LWS full grating, and CAM CVF. We are in the 
process of analyzing these data (see e.g. Cenicharo et al. 1999). 
Unlike HH 7-11 or Cep E, the HH 1-2 objects are {\it high excitation} 
objects (Fig. 6), i.~e. with shock velocities of $\sim 100$ \kms, enough to 
produce high ionization [O~III] $\lambda$5007 emission.
Unfortunately the observations of HH 1 have been ``contaminated'' by the 
bright Cohen-Schwartz source that lies nearby along the outflow. The H$_2$ 
pure rotational S(7) through S(1) were observed by SWS, as well as 
[Si~II] 34.8\mum. CAM CVF detected the H$_2$ (0,0) S(2) - S(7) lines, but also
in some knots (HH 2H) [Ne~II] 12.8\mum~and [Ne~III] 15\mum, indicating 
v$_{shock}> 60$ \kms~(Fig. 7). The LWS spectra was taken at 3 positions (HH 1,
2 and VLA 1 source) and no molecular lines of CO, H$_2$O, and OH were detected.
The fine structure atomic lines of [O~I] 63\mum~and [C~II] 158\mum are
clearly seen and with a similar brightness. This suggests a strong 
photodissociation component due to the surrounding UV field that affects the 
strength of the [C~II] line.

\begin{figure}
  \begin{center}
    \leavevmode
     \includegraphics[height=8.0cm,width=9.0cm]{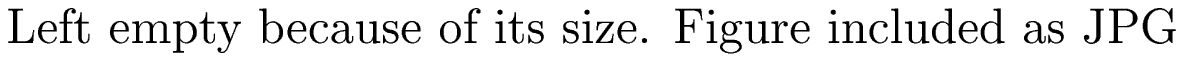}
    \caption{ISOCAM CVF (4-17\mum) image of HH 2 \& VLA 1 region
     [from Cernicharo et al. 1998]. The emission of HH 2 contains
      H$_2$ (0,0) rotational lines as well as [Ne~II] 12.8\mum~\&
      [Ne~III] 14.5\mum. Dust emission from the jet is also detected.}
    \label{fig:anc_f07.ps}
  \end{center}
\end{figure}

\section{Class 0/I sources}
\label{sources}

 A very nice result from the ISO observations came from the CAM CVF 
observations towards Class 0 sources. A Class 0 source, by definition, is a 
true protostar with the peak of its spectral energy distribution at sub-mm or 
FIR wavelengths. They are deeply embedded inside an ``envelope'', detected in 
many cases thanks to their powerful bipolar outflows. Our first CAM 
observations of Cep E detected, {\it at mid infrared wavelengths}, 
the driving source (Noriega-Crespo et al. 1998),
presumably a Class 0 source. The CVF observations confirmed this detection and
displayed a series of spectral absorption features between 5-17\mum~
(CH$_4$ at $\sim 7.5$, Silicates at 9-10\mum, and CO$_2$ at 14-15\mum) 
closer to those expected from a more evolved Class I source.
We found a similar result for the VLA 1 source in the HH 1-2 outflow and 
other well known class 0 sources, like NGC1333-IRAS2 or L1448-N 
(Cernicharo et al. 2000). The difference with respect to a source like 
IRAS230111+6126, is that the absorption features are so deep, that what is 
left in the mid-IR SED are a series of emission windows 
at $\sim 5.3, 6.6~\& 7.5$\mum~(see Fig. 8). In the case of VLA 1, 
for instance, we can reproduced the mid-IR SED with a object of 4AU in size 
at 700K and an extinction of A$_V$ = 80$-$100 magnitudes (soft line in Fig. 8).

\begin{figure}
  \begin{center}
    \leavevmode
     \includegraphics[height=7cm,width=5cm]{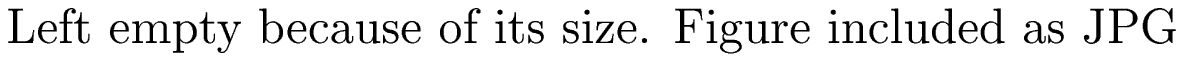}
    \caption{A sample of Class 0/I sources showing their similarity
            at mid-IR wavelengths [from Cernicharo et al. 2000]}
    \label{fig:anc_f08.ps}
  \end{center}
\end{figure}

\section{General Trends: Examples}
\label{trends}

\subsection{Cooling from C-shocks}
\label{cool}

When the ISO data archive became public in December 1998, 
it was clear that one could have access to a large sample of objects and 
able to ask more global questions about the behavior and energetics 
of outflows, jets and their sources. An example of this
approach was carried out by Paolo Saraceno and its group based on
LWS observations of Class 0$-$III sources (Saraceno et al. 1997).
In Fig. 9 we used a similar idea (see e.g Spignolio et al 2000) to
determine the shock velocity in a series of outflows, based on their
H$_2$O, [O~I] 63\mum~and CO cooling when compared with the 
predictions from C-type shocks (Draine et al. 1983; 
Kaufmann \& Neufeld 1996). At first approximation the cooling from these 
species defines a narrow range of shock velocities ($10-20$\kms) and 
densities (log n(H$_2$)= $4.5-5.5$), with
perhaps a couple of exceptions (IC1396N \& IRAS 16293). If we approximate the
working surfaces of the jets/outflows by  bowshocks, then the above result 
suggests that the molecular emission arises away from the stagnation region,
further ``downstream'' and along the bow shock wings. A result compatible
with the numerical hydrodynamics simulations of working surfaces, 
with parallel integrated chemistry (Williams 2001; Lim et al. 1999).

\begin{figure}
  \begin{center}
    \leavevmode
     \includegraphics[height=8cm,width=7cm]{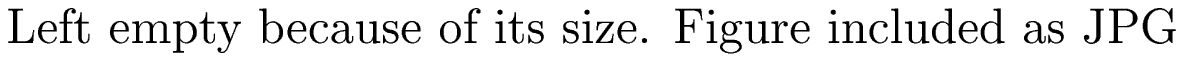}
    \caption{The molecular cooling at FIR from several outflows, observed
     with LWS, compared with the predictions from C-shock models.}
    \label{fig:anc_f09.ps}
  \end{center}
\end{figure}

\subsection{H$_2$O from Circumstellar Envelopes}
\label{envelope}

LWS observations have been used recently as well (Ceccarelli et al. 1999),
to study the correlation between H$_2$O thermal emission arising from 
proto-stellar sources and their 1.3mm continuum fluxes or SiO millimetric 
emission. The idea is to try to distinguish if the H$_2$O emission comes from
the outflows (SiO as a tracer) or closer to the source (an envelope, with 1.3mm
flux as a tracer). Although the results are far from conclusive (see e.~g.
Neufeld et al. 2000), from a sample of 7 YSOs, 5 seem to correlate quite 
well with their 1.3mm fluxes (Fig. 10), but not with the SiO emission.
The suggested explanation is that the H$_2$O lines originate
from the warm inner region of infalling envelope, with accretion rates
of a few $10^5$\msol/yr.

\begin{figure}
  \begin{center}
    \leavevmode
     \includegraphics[height=8cm,width=7cm]{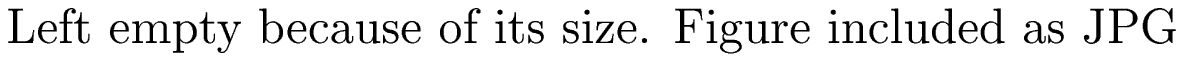}
    \caption{H$_2$O thermal emission from a sample of YSOs
     compared with their 1.3mm fluxes [from Ceccarelli et al. 1999]}
    \label{fig:anc_f10.ps}
  \end{center}
\end{figure}

\section{Summary}
\label{summa}

 Astronomically ISO was a very successful mission, and for star forming 
regions, stellar jets, outflows and their sources, it meant hundreds of new
observations at mid and far infrared wavelengths. I have shown a small sample
of the spectroscopic results on some prototype outflows, as HH 1-2, HH 7-11
and Cep E, that have illustrated the different ISO observational modes, 
and the physical conditions that can be derived from them. It is clear that
molecular cooling at FIR wavelengths is an important component of the
overall energy budget of outflows and jets. One of the nice surprises from
ISO has been the possibility to observe deeply embedded sources at 
mid-IR wavelengths, since this opens a door for ground based observations 
with large telescopes of some of the less known phases of the star formation
process.

\acknowledgments

It is a pleasure to thank Alex Raga and Luc Binette for their warm welcome and 
the great organization of this meeting. A pleasure to thank my collaborators:
B. Ali, S. Cabrit, C. Ceccarelli, J. Cernicharo, S. Molinari, 
A. Moro-Mart{\'{i}}n, 
P. Saraceno, A. Sargent, and L. Testi, as well.  Thanks to the SOC for the 
invitation. Last, but not least, to Jorge Cant\'o, Adam Frank, 
Jos\'e Alberto Lop\'ez, Alex Raga, and Jose Maria Torrelles for enlightened 
conversations. {\it Gracias} to A. Moro-Mart{\'{i}}n for her careful
reading of this manuscript.

%\vspace*{4cm}

\end{document}